\documentclass[reprint,prl,twocolumn,superscriptaddress,showpacs,nofootinbib,floatfix,preprintnumbers]{revtex4}
\usepackage{axodraw}
\usepackage{graphicx,bm}

\def\beq {\begin{equation}}
\def\eeq {\end{equation}}
\def\bea {\begin{eqnarray}}
\def\eea {\end{eqnarray}}

\newcommand{\br}{\begin{eqnarray}}
\newcommand{\er}{\end{eqnarray}}
\newcommand{\be}{\begin{equation}}
\newcommand{\ee}{\end{equation}}

\usepackage{color}
\usepackage{hyperref} 
\hypersetup{linktocpage=true}
\usepackage[all]{hypcap}
\hypersetup{
   bookmarks=true,         % show bookmarks bar?
   unicode=false,          % non-Latin characters in AcrobatÍs bookmarks
   pdftoolbar=true,        % show AcrobatÍs toolbar?
   pdfmenubar=true,        % show AcrobatÍs menu?
   pdffitwindow=false,     % window fit to page when opened
   pdfstartview={FitH},    % fits the width of the page to the window
   pdftitle={My title},    % title
   pdfauthor={Author},     % author
   pdfsubject={Subject},   % subject of the document
   pdfcreator={Creator},   % creator of the document
   pdfproducer={Producer}, % producer of the document
   pdfkeywords={keyword1} {key2} {key3}, % list of keywords
   pdfnewwindow=true,      % links in new window
   colorlinks=true,       % false: boxed links; true: colored links
   linkcolor=blue,          % color of internal links
   citecolor=magenta,        % color of links to bibliography
   filecolor=magenta,      % color of file links
   urlcolor=cyan           % color of external links
}
\usepackage{fancyhdr}
\begin{document}
\preprint{DAMTP-2014-50, IPPP/14/78, DCPT/14/156} 
\title{Explaining a CMS $eejj$ Excess With
  $\mathcal{R}-$parity Violating Supersymmetry and Implications for
  Neutrinoless Double Beta Decay}

\author{Ben Allanach} \affiliation{DAMTP, CMS, Wilberforce Road, University of
  Cambridge, Cambridge, CB3 0WA, United Kingdom}

\author{Sanjoy Biswas } \affiliation{Dipart. di Fisica, Universit\`a di Roma
  “La Sapienza”, Piazzale Aldo Moro 2, I-00185 Rome, Italy}

\vspace*{0.3cm}

\author{Subhadeep Mondal } \affiliation{Department of Theoretical Physics,
  Indian Association for the Cultivation of Science, 2A $\&$ 2B Raja
  S.C. Mullick Road, Kolkata 700032, India}

\vspace*{0.3cm} \author{Manimala Mitra} \affiliation{Institute for Particle
  Physics Phenomenology, Department of Physics, Durham University, Durham DH1
  3LE, United Kingdom}

\begin{abstract}
A recent CMS search for the right handed gauge boson $W_R$ reports an
 interesting deviation  
from the Standard Model. The search has been conducted in the  
$eejj$  channel and has shown a 2.8$\sigma$ excess around $m_{eejj} \sim 2$
 TeV. In this work, we explain  the  
reported CMS excess with R-parity violating supersymmetry (SUSY). We consider
resonant selectron and sneutrino  production,  
followed by the three body decays  of 
the neutralino and chargino  via an $\mathcal{R}-$parity violating
coupling. We  
fit the excess for slepton masses around 2 TeV. The scenario
can further be tested in neutrinoless double beta  
decay ($0\nu \beta \beta$) experiments.
GERDA Phase-II will probe a significant portion of the good-fit parameter
space. 
\end{abstract}
% \pacs{}
\maketitle

The recent CMS search 
for a hypothetical $W_R$ gauge boson in the Left-Right model reports an
 intriguing deviation  
from the Standard Model in the $eejj$ channel.  The CMS search uses $pp$
collision data at the Large Hadron 
Collider (LHC) and a centre of mass energy of $8$ TeV and $19.7 \rm{fb}^{-1}$
of integrated luminosity. The invariant mass distribution $M_{eejj}$ shows an
excess around $M_{eejj}\sim 2$ TeV, with a local CERN CL$_s$ significance of
2.8$\sigma$ 
\cite{Khachatryan:2014dka}.  In the $1.8$ TeV$< m_{eejj}<2.2$ TeV bin, CMS
reported 14 events on an expected background of 4.0$\pm$1.0.
However, no significant deviation was 
observed in the $\mu\mu jj$ channel.  
This excess is not significant enough to claim a
discovery. However, it is timely
before the next LHC run (Run II) to explain it with a concrete
model of new 
physics such that further tests can be applied and analysis strategies can be
set for Run II.

There have been a few attempts to explain the CMS
excess with  different models. Coloron-assisted
leptoquarks were proposed in Ref.~\cite{Bai:2014xba}. 
The $W_R$ excess was interpreted in GUT models in
Refs.~\cite{Deppisch:2014qpa,Heikinheimo:2014tba}. In
Ref.~\cite{Dobrescu:2014esa}, pair production of vector-like leptons
was proposed via $W'/Z'$ vector bosons.
Ref.~\cite{Aguilar-Saavedra:2014ola} performed a detailed analysis (including a
general flavor structure) of $W'/Z'$ interpretations of the $W_R$ search data.

In this letter, we propose a different hypothesis for a new physics explanation
of the $W_R$ search excess in terms of the $\mathcal{R}$-parity violating
(RPV) 
minimal supersymmetric 
model (MSSM).  $\mathcal{R}$-parity
is a multiplicative discrete symmetry
defined as 
$\mathcal{R}=(-1)^{3(B-L)+2S}$, where $B$ and $L$ correspond to baryon and
lepton number, and $S$ is spin.  
In particular, we show that
RPV with a non-zero $\lambda'_{111}$ coupling
can fit the 
CMS excess \cite{Khachatryan:2014dka,CMS1} via resonant slepton
production (with a slepton mass of around 2 TeV) in $pp$ collisions.  The
slepton 
then 
subsequently decays to a charged lepton and neutralino, followed by the RPV decay modes of the neutralino via the $\lambda'_{111}$ coupling, producing an excess of
events in the $eejj$ channel, as depicted in Fig.~\ref{fig:FDsel}. The same 
signature in the $eejj$ channel can also be obtained from the resonant 
production of a sneutrino, followed by the R-parity violating decays 
of charginos, as shown in Fig.~\ref{fig:FDsnu}. 

%%%%%%%%%%%%%%%%%%%%%%%%%%%% ...Figure-1
\begin{figure}[ht]
  % \begin{center}
  %   \vspace*{-2.2cm}
  \begin{picture}(180,60)(0,0)
    \ArrowLine(0,0)(30,30)
    \ArrowLine(0,60)(30,30)
    \DashLine(30,30)(60,30){5}
    \ArrowLine(60,30)(90,60)    
    \ArrowLine(60,30)(90,0)    
    \ArrowLine(120,0)(90,0)
    \ArrowLine(120,0)(120,30)
    \DashLine(120,0)(150,0){5}    
    \ArrowLine(150,0)(150,30)
    \ArrowLine(150,0)(180,0)
    \Text(18,30)[c]{$\lambda'_{111}$}
    \Text(165,10)[c]{$\lambda'_{111}$}
    \Text(0,53)[c]{$\bar u$}    
    \Text(0,7)[c]{$d$} 
    \Text(45,35)[c]{$\tilde e_L$}       
    \Text(90,53)[c]{$e$}       
    \Text(90,10)[c]{$\chi_1^0$}       
    \Text(115,25)[c]{$u$}       
    \Text(145,25)[c]{$e$}       
    \Text(180,7)[c]{$d$}       
    \Text(135,7)[c]{$\tilde u_L$}       
  \end{picture}
  % \includegraphics[width=7.5cm]{anc/FD-selectron.eps}
  % \includegraphics[width=7.5cm]{anc/FD-sneutrino.eps}
  % \centerline{\epsfig{file=lhc_.eps,width=9.0cm,height=7.5cm,angle=-0}}
  \caption{Feynman diagram for single selectron production leading to 
$eejj$ signal at the LHC.}
  \label{fig:FDsel}
  % \end{center}
\end{figure}
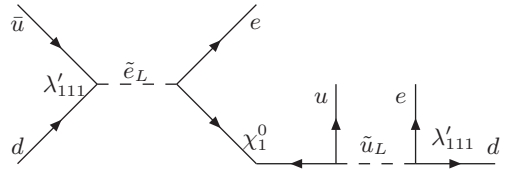
\begin{figure}[ht]
  % \begin{center}
  %   \vspace*{-2.2cm}
  \begin{picture}(180,60)(0,0)
    \ArrowLine(0,0)(30,30)
    \ArrowLine(30,30)(0,60)
    \DashLine(30,30)(60,30){5}
    \ArrowLine(60,30)(90,60)    
    \ArrowLine(60,30)(90,0)    
    \ArrowLine(90,0)(120,0)
    \ArrowLine(120,30)(120,0)
    \DashLine(120,0)(150,0){5}    
    \ArrowLine(150,30)(150,0)
    \ArrowLine(150,0)(180,0)
    \Text(18,30)[c]{$\lambda'_{111}$}
    \Text(165,10)[c]{$\lambda'_{111}$}
    \Text(0,53)[c]{$ d^c$}    
    \Text(0,7)[c]{$d$} 
    \Text(45,35)[c]{$\tilde \nu_L$}       
    \Text(90,53)[c]{$e$}       
    \Text(90,10)[c]{$\chi_1^+$}       
    \Text(115,25)[c]{$e^c$}       
    \Text(145,25)[c]{$d^c$}       
    \Text(180,7)[c]{$d$}       
    \Text(135,7)[c]{$\tilde \nu_L$}       
  \end{picture}
  % \includegraphics[width=7.5cm]{anc/FD-selectron.eps}
  % \includegraphics[width=7.5cm]{anc/FD-sneutrino.eps}
  % \centerline{\epsfig{file=lhc_.eps,width=9.0cm,height=7.5cm,angle=-0}}
  \caption{Feynman diagram for single sneutrino production leading to 
$eejj$ signal at the LHC.}
  \label{fig:FDsnu}
  % \end{center}
\end{figure}
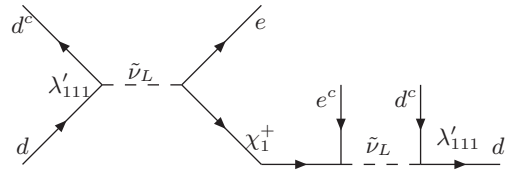
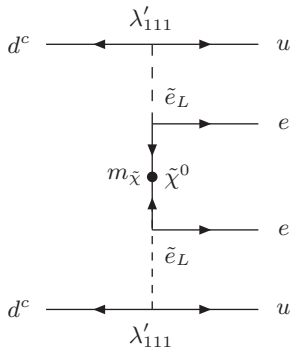
\begin{figure}[b]
\vspace*{1.12cm}
  \begin{picture}(180,20)(0,0)
\ArrowLine(90,40)(50,40)
\ArrowLine(90,40)(130,40)
\DashLine(90, 10)(90,40){4}%{4}
\ArrowLine(90,10)(90,-10)
\ArrowLine(90,-30)(90,-10)
\ArrowLine(90,10)(130,10)
\ArrowLine(90,-30)(130,-30)
\DashLine(90, -20)(90,-60){3}%{3}
\ArrowLine(90,-60)(50,-60)
\ArrowLine(90,-60)(130,-60)
\Vertex(90,-10){2}
\Text(40,40)[]{$d^c$}
\Text(140,40)[]{$u$}
%\Text(150,30)[]{$\tilde{\lambda}^+_1$}
\Text(40,-60)[]{$d^c$}
\Text(100,-10)[]{$\tilde{\chi}^0$}
\Text(100,20)[]{$\tilde{e}_L$}
\Text(100,-40)[]{$\tilde{e}_L$}
\Text(140,-60)[]{$u$}
\Text(90,-70)[]{$\lambda'_{111}$}
\Text(90,50)[]{$\lambda'_{111}$}
\Text(140,-30)[]{$e$}
\Text(140,10)[]{$e$}
\Text(80,-10)[]{$m_{\tilde{\chi}}$}
%\caption{}
\end{picture}
\label{0nu2beta}
\vspace*{2.70cm}
\caption{Sample Feynman diagram for $0\nu \beta \beta$,  corresponding to 
 selectron and neutralino contribution. There are several other diagrams from
  gluino and  squark mediation, that contribute to the half-life $T^{0\nu}_{1/2}$ \cite{Allanach:2009xx}. In our analysis of $0\nu \beta \beta$, we consider all possible contributions via the $\lambda'_{111}$ coupling \cite{hirsch, Allanach:2009xx}.  }
\end{figure}

%%%%%%%%%%%%%%%%%%%%%%%%%%%%%%%%%%%%%%%%%%%%%%%%%%%%%%%%%
%This in turn can strongly constrain the allowed
%values of $\lambda'_{111}$ coupling. 
The RPV superpotential with the $\lambda'_{111}$ term is 
%$\lambda'_{111}$ term in the $\mathcal{R}-$parity violating superpotential is
\begin{equation}
  W_{\not{R}}= \lambda'_{111} LQd^c. 
  \label{eqrpv}
\end{equation}
This induces the following Lagrangian terms,
\begin{equation}
  \mathcal{L}=-\lambda'_{111} \tilde{e} u d^c-\lambda'_{111} \tilde{u} e d^c+
%- \lambda'_{111} \tilde{d^*} e u+ 
\lambda'_{111} \tilde{d} \nu_e d^c+ \tilde{\nu}_e d d^c+\ldots\\ \label{eq:L}
\end{equation}
The MSSM with $\lambda'_{111}$ is constrained by empirical data on
charge current universality, $e$--$\mu$--$\tau$
universality, atomic parity violation etc~\cite{Allanach:1999ic}.
 In addition, the model contributes 
to lepton number violating 
neutrinoless double beta decay ($0\nu \beta \beta$) \cite{0nu2beta-old,
  ms0nu2beta,  hirsch}, as shown in Fig.~\ref{0nu2beta}. $0\nu\beta\beta$ is  
 not permitted in the SM because of lepton number conservation.
{The present bound on the half-life  of $^{76}\rm{Ge}$ isotope is
  $T^{0\nu}_{1/2}>2.1 \times 10^{25}\, \rm{yrs}$ at 90$\%$ confidence
  level (CL) from GERDA~\cite{gerda},
  while the 90$\%$ CL combined bound on the half-life  from  previous
  experiments 
  is $T^{0\nu}_{1/2}>3.0 \times 
  10^{25}\, \rm{yrs}$~\cite{gerda}. The future  
$0\nu \beta \beta$ experiment  GERDA Phase-II will be commissioned soon and
is expected to  
improve the half-life sensitivity to  $T^{0\nu}_{1/2} \sim 2\times 10^{26} \,
\rm{yrs}$~\cite{gerdafuture}.}  
A positive signal in $0\nu \beta \beta$ experiments is likely to be
interpreted in terms of a Majorana nature of the light neutrinos, 
but instead it could be  in part, or dominantly, due to RPV SUSY. 

{ The most 
stringent bound on the $\lambda'_{111}$ coupling is 
approximately  proportional to 
$m_{\tilde l_R}^2 (m_{\chi_1^0})^{1/2}$ that comes from $0\nu\beta\beta$ (see.
Ref.~\cite{Allanach:2009xx}) and is  shown in Table \ref{tab:bounds}. We shall calculate this bound more precisely, i.e. at the diagramatic level.
While the bounds in the table are for 100 GeV sparticles, they become greatly
weakened for the heavier sparticles that  we shall consider. In addition, we also show the other bounds from universality \cite{Reviews}. }

The $\lambda'_{111}$ coupling in Eq.~\ref{eq:L} can lead to  single slepton
production at hadron colliders, as %.  
%The resonant
%slepton and sneutrino production has been first studied
first studied in \cite{Dimopoulos:1988fr} and subsequently in 
\cite{Hewett:1998fu, Dreiner:1998gz, Dreiner:2000vf, Dreiner:2000qf, Dreiner:2012np, Allanach:2003wz, Allanach:2009xx, Allanach:2009iv}. For a slepton of
mass around the CMS excess (2.1 TeV) and $0.03<\lambda'_{111}<0.5$ the
production cross-section varies from less than 1 fb to as high as 130
fb \cite{Dreiner:2000vf}.  
As pointed out in Ref.~\cite{Allanach:2009iv}, one can marry resonant
slepton search data from the LHC with the predicted $0\nu\beta\beta$ rate in
order to provide further tests and interpretations. 
The $\lambda'_{111}$ coupling also leads to the resonant  
production of sneutrinos, as shown in Fig.~\ref{fig:FDsnu}. 
The decay mode of the sneutrino leading to the $eejj$ signal is:
$pp \rightarrow \tilde \nu_e/ \tilde \nu^*_e \rightarrow e^- \chi_1^+/ e^+
\chi_1^- \rightarrow 
e^+e^-jj$. 
It is our aim to see if resonant selectron and sneutrino production can fit
the CMS $W_R$ excess while evading other experimental constraints. 

%, 
%or $pp \rightarrow \tilde \nu_e \rightarrow e^+ \chi_1^- \rightarrow e^+e^-jj$.
\begin{table}
  \centering
  \begin{tabular}{|c|c|}\hline
    Bound & Origin  \\ \hline
    0.05$^*$  & Lepton flavour universality of $\pi^\pm$ decay\\
    0.02  & Charge current universality \\
    { 5$\times 10^{-4}$}   & $0\nu\beta\beta$ \\ \hline
  \end{tabular}
  \caption{Upper bounds on the  $\lambda'_{111}$
    coupling, assuming all sparticle masses to be 100 GeV. The bounds are at
    95$\%$ confidence level (CL), except for that marked $^*$, which is at the
    68$\%$ CL.} 
  \label{tab:bounds}
\end{table}
 In this letter, we follow  a bottom-up phenomenological 
approach. We fix any sparticles which
are not relevant for our hypothesised signals  to be heavy enough
not to be produced at the LHC. Otherwise, we  
fix the first generation left-handed slepton mass to be 
2.1 TeV,
the lightest neutralino mass varies from 400 GeV up to 1 TeV and all other
sparticles are above the TeV scale. 
The squarks are fixed at 2 TeV
masses (the $0\nu\beta\beta$ rate we predict below depends somewhat on this
assumption due to additional diagrams to Fig.~\ref{0nu2beta} involving squarks).
In addition, we 
set other RPV couplings to zero, allowing us to focus purely  on the effects of
$\lambda'_{111}$. 

The phenomenology  is model dependent. We have
considered the following representative scenarios:
\begin{itemize}
\item {\bf S1:} $M_1 < M_2=M_1+200 < \mu$, {i.e.}, the LSP is mostly bino-like
  with a small wino-component.  In this case the slepton has a substantial
  branching ratio of decaying to the second lightest neutralino or lightest
  chargino.
\item {\bf S2:} $M_1 < \mu < M_2 $, the LSP is still dominated by the
  bino-component, with a heavy intermediate higgsino mass and an even heavier
  wino mass ($> 1$ TeV). This case is interesting because it increases the
  branching ratio of slepton decaying into the lightest neutralino and a
  lepton.
\item {\bf S3:} $M_2 << M_1 \simeq \mu$, {i.e.}, the LSP is dominantly
  wino-like. In this case,  the slepton also decay to lighter chargino and a
  neutrino with a substantial branching fraction. In this scenario, both the
  lighter chargino and lightest neutralino decay via $\lambda'_{111}$. Hence, the
  lepton and jet multiplicities in the final state are enhanced compared to 
  {\bf S1}, {\bf S2}.
\end{itemize}

\begin{figure}[!]
  \includegraphics[width=7.0cm]{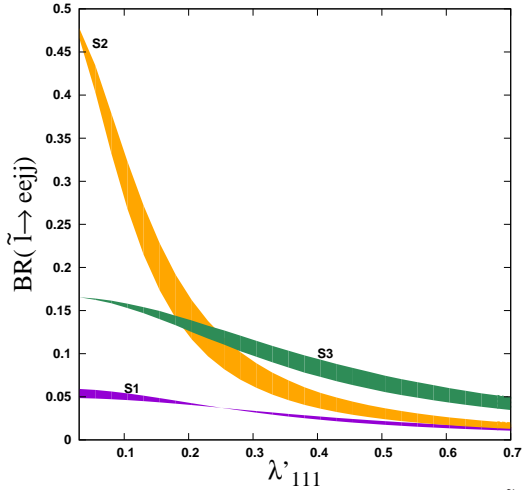}
\vspace*{-0.5cm}  
\caption{{The effective branching ratio of the decay $\tilde{\ell} \rightarrow
 eejj$ in ${\bf S1}-{\bf S3}$
for possible choices of $\lambda'_{111}$ coupling. The small color bands indicate the variation of the 
branching ratio with $m_{\tilde{\chi}^0_1} (400 ~{\rm GeV} - 1.0 ~{\rm TeV})$
for a given $\lambda'_{111}$.  
}}
  \label{fig:br}
\end{figure}

%Other scenarios, for example where the LSP is purely Higgsino-like, are not
%interesting for this study. 
Depending on the nature of the lightest
neutralino and the value of the $\lambda'_{111}$ coupling, the branching ratio
changes considerably \cite{Dreiner:2012np}. 
% For example, assuming the neutralino to be
% purely bino-like and $0.03<\lambda'_{111}<0.5$ the branching ratio
% Br($\tilde{e} \rightarrow e {\chi}^0_1$) varies from 90\% to 5\%. 
We show the effective branching fraction  Br($\tilde{l} \to eejj$)  for our
various 
scenarios in Fig.~\ref{fig:br}.
We note that a higher value of $\lambda'_{111}$ leads to stringent
limits  from di-jet resonance searches.  
We take into account the constraint from CMS di-jet resonance
search \cite{Chatrchyan:2013qha}.  
%  We have imposed the constraint
 % coming from the CMS di-jet resonance search [Ref.]. 
The 
  limit on the cross-section for a di-jet resonance around 2.1 TeV is $< 45$
  fb. This 
  in turn gives a bound on the product ${\lambda'_{111}}^2 \times
  Br(\tilde{e}/\tilde{\nu} \rightarrow j j)$ \cite{Dreiner:2012np}.

\begin{figure}[t]
  \includegraphics[width=7.0cm]{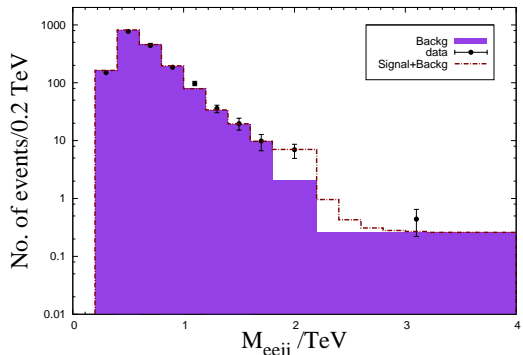}
  \vspace*{-0.5cm}  
  \caption{{A comparison of the data, signal and background
      $M_{eejj}$ distributions after imposing cuts as done in the analysis of
      the $W_R$ search. The signal point corresponds to
      $\lambda'_{111}=0.105$ and $m_{\tilde{\chi}^0_1}=532$ GeV ({\bf
        S3}). The data and SM       backgrounds are taken from
      \cite{Khachatryan:2014dka}.
}}
  \label{fig:Meejj}
\end{figure}

We simulate first generation resonant slepton production in $pp$
collisions at a centre of mass energy $\sqrt{s}=8$ TeV using CalcHEP (v3.4.2)
\cite{Belyaev:2012qa},
and the subsequent decay, showering and hadronization effects have been
performed 
by PYTHIA (v6.4) \cite{Sjostrand:2006za}. We use SARAH-v4.0.1 \cite{sarah} and SPheno-v3.2.4 \cite{spheno}
for the model implementation and to compute  branching ratios. We approximate
the next-to-leading order QCD corrections by multiplying the tree-level
production cross section 
with a $K-$factor of 1.34~\cite{Dreiner:2012np}. We use CTEQ6L parton
distribution
functions~\cite{Kretzer:2003it}  with factorization and  renormalization
scales set at 
the slepton mass $\tilde{m}_L$. To take into account 
detector resolution effects, we  also use various resolution functions
parameterized as in \cite{Chatrchyan:2011ds} for the final state objects. 

%{\bf The NLO correction...We have taken into account appropriate K-factor=1.34
 % [Ref.]}

The final state studied in \cite{Khachatryan:2014dka}, contains exactly two
isolated leptons and at least two jets ($2\ell+\geq 2j$). Basic object
definitions for the leptons and jets together with the following final
selection cuts, as 
outlined in \cite{Khachatryan:2014dka}, have been imposed:
\begin{itemize}
\item Invariant mass of the lepton pair, $M_{\ell\ell}> 200$ GeV.
\item Invariant mass of the leptons and two hardest jets  $M_{\ell\ell jj}>
  600$ GeV. 
\end{itemize}

We assume a truncated Gaussian for the prior probability density function
(PDF) of $\bar b \pm \sigma_b$ background events: 
\begin{equation}
p(b | \bar b,\ \sigma_b) = \left\{ 
\begin{array}{lr}
B e^{-(b-\bar b)^2/(2 \sigma_b^2)}& \forall b > 0 \\
0& \forall b \leq 0
\end{array} \right.
\end{equation}
where $B$ is a normalisation factor that makes the distribution integrate to 1.
We marginalise the Poissonian probability of measuring $n$ events over $b$
in
order to obtain confidence limits:
\begin{equation}
P(n|n_{exp},\ \bar b,\ \sigma_b) = \int_0^\infty db\ p(b | \bar b,\ \sigma_b) \frac{e^{-n_{exp}} n_{exp}^{n}}{n!},
\end{equation}
where $n_{exp}$ is the number of expected events.
The CL of $n_{obs}$ observed events is then $P(n\leq n_{obs})$.
Calculated in this way, the local significance of the
$1.8<M_{eejj}/\text{TeV}<2.2$ bin is 
3.6$\sigma$\footnote{The $CL_s$ method employed by CMS yields 3.2$\sigma$ for
  these assumed statistics. The discrepancy between this number and the quoted
2.8$\sigma$ comes from separate systematic errors on the different background
components, which we do not have access to here.}. The two-sided 95$\%$CL
bound on the number of signal events in this bin is $s\in[4.1-19.7]$.

We present our results in  Table.~\ref{tab:events}  and in
Fig.~\ref{fig:Meejj} for a typical {\bf S3} scenario.  
In Table.~\ref{tab:events}, we show the event rate assuming an integrated
luminosity of 19.7 fb$^{-1}$  and the
corresponding experimental data and SM backgrounds.
\begin{table}[!]
  \small
  % \begin{center}
  %   \tabulinesep=1.2mm
  \begin{tabular}{|l|c|c|c|} 
    \hline 
    Cut & {Signal} & Background & Data \\
    \hline
    $2e+\geq 2j$  &  12.7       &  34154  &   34506   \\
    $M_{ee}> 200$ GeV  &  12.6      &   1747  &   1717  \\
    $M_{eejj}> 600$ GeV  &    12.6       &   783$\pm$51  &  817   \\
    1.8 TeV$< M_{eejj} <$ 2.2 TeV  &  10      &  4.0$\pm$1.0   &   14   \\
    \hline
  \end{tabular}
  \caption{Number of events from signal, backgrounds and
 reconstructed data  
 after  
    successive application of the selection cuts at 19.7 fb$^{-1}$ integrated
    luminosity and 
    8 TeV center of mass energy for scenario {\bf S3} assuming $\lambda'_{111}=0.105$ and
 $m_{\tilde{\chi}^0_1}=532$ GeV. The data and SM backgrounds are taken from
 Ref.~\cite{Khachatryan:2014dka}.} 
  \label{tab:events}
  % \end{center}
\end{table} 
%%%%%%%%%%%%%%%%%%%%%%%%%%%%%%%%%%%%%%%%%%%%%%%%%%%%%%%%%%%%%%%%%%%%%
%
%%%%%%%%%%%%%%%%%%%%%%%%%%%% ...Figure-3a
\begin{figure}[!]
  \includegraphics[width=8.5cm]{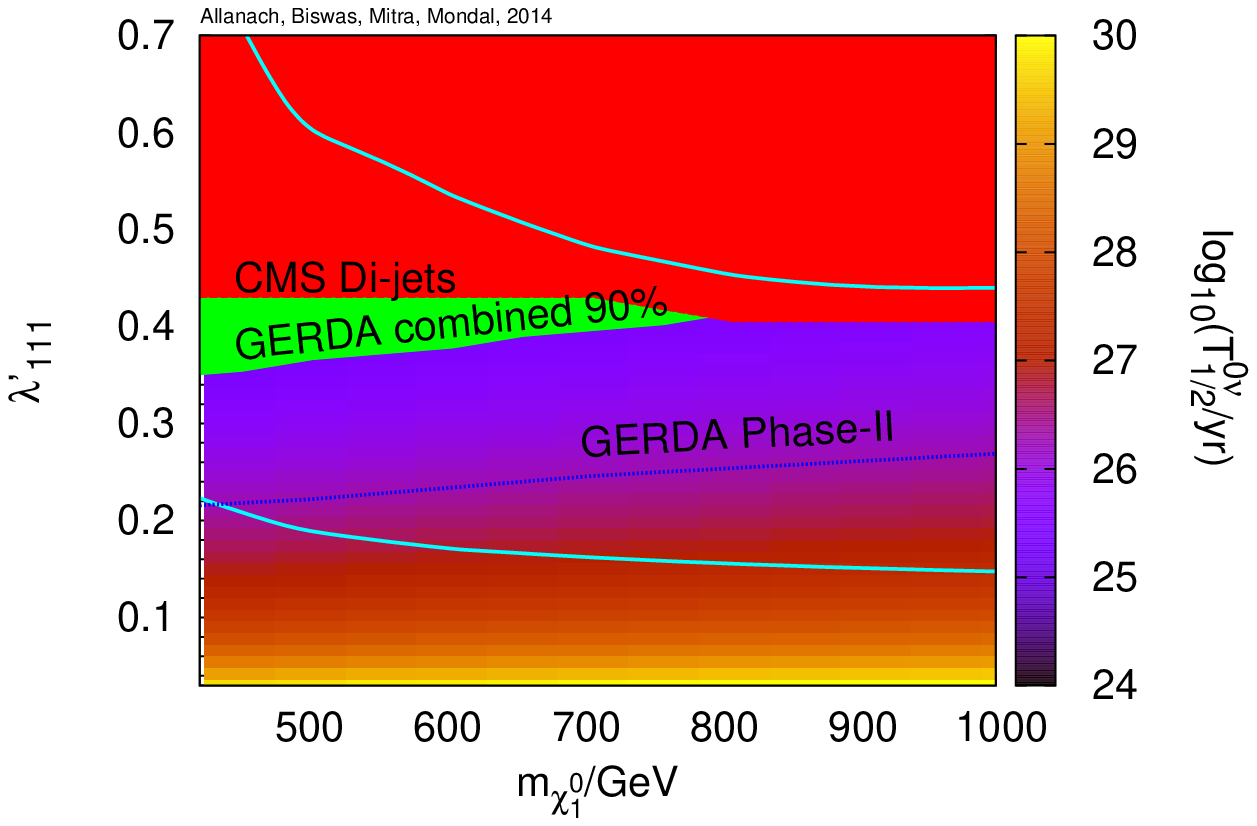}
  \includegraphics[width=8.5cm]{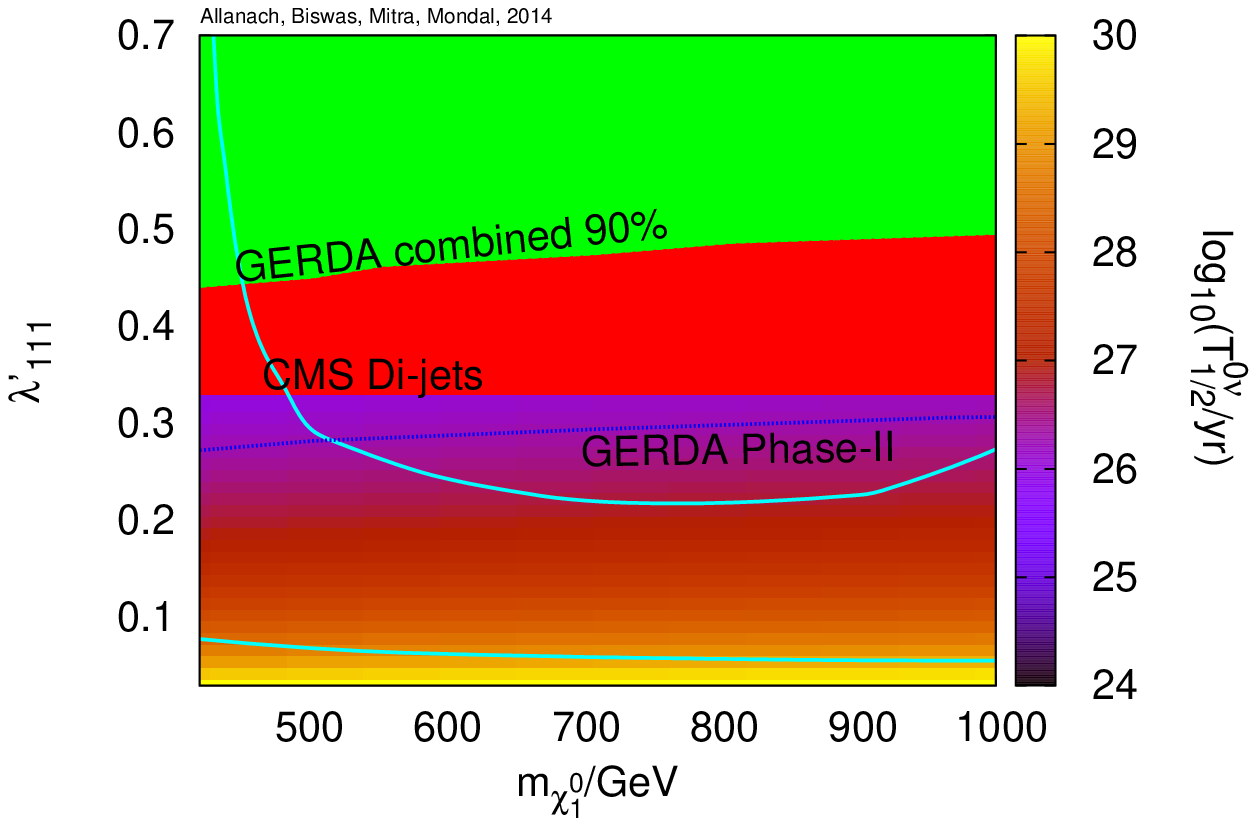}
  \includegraphics[width=8.5cm]{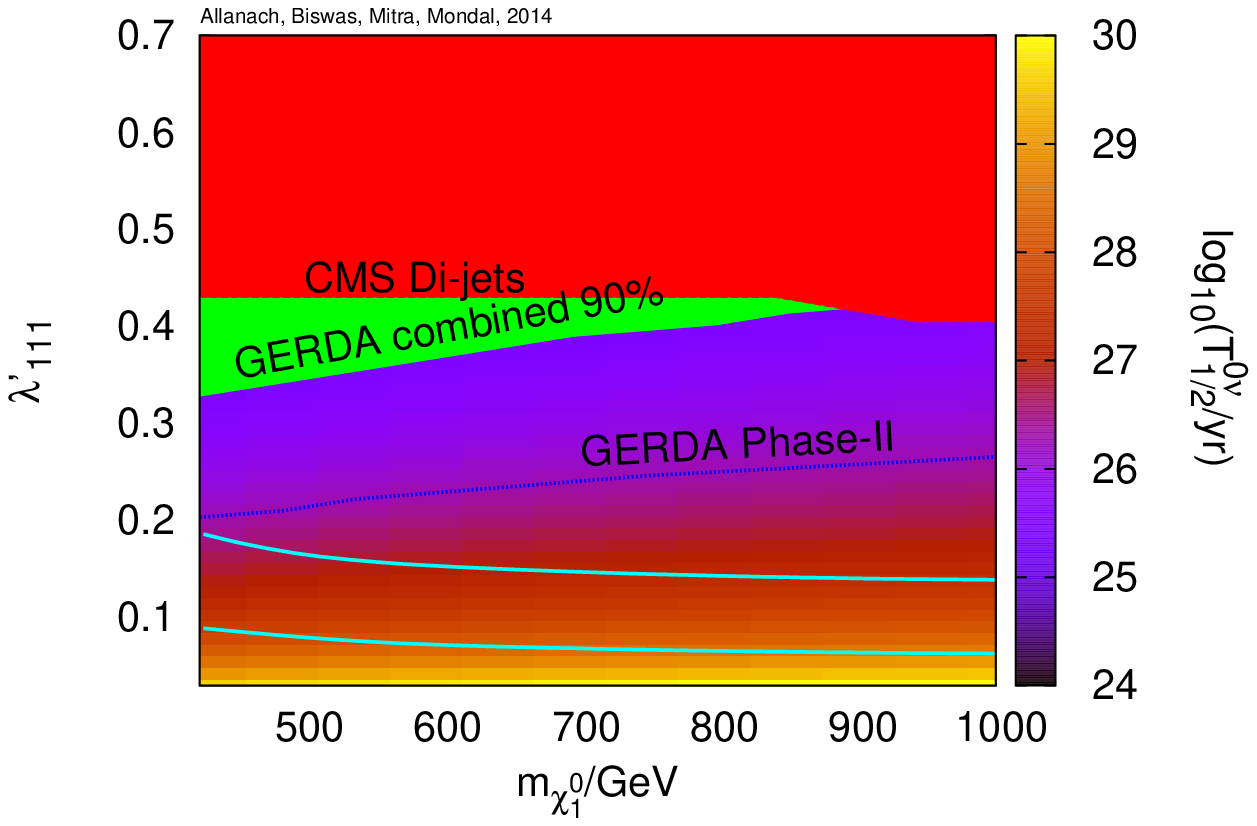}
\vspace*{-0.5cm}  
\caption{A scan in the $\lambda'_{111}$ coupling and the
    neutralino mass plane assuming 2.1 TeV slepton mass and scenario (top)
    {\bf S1}, (middle) {\bf S2} and (bottom) {\bf S3}. The color gradient
    represents  
    the half-life $T^{0\nu}_{1/2}$ of $0\nu \beta \beta$ process, 
    where the nuclear matrix uncertainty has been adopted from 
    \cite{Allanach:2009xx}. 
    The
    region between the light curves fit data from the  bin 
    $1.8$ TeV$< m_{eejj}<2.2$ at the 95$\%$ CL level.
    We show regions excluded at
    95$\%$CL by the CMS di-jet resonance search~\cite{Chatrchyan:2013qha} and
    the 90$\%$ CL current combined constraints coming from $0\nu\beta\beta$ 
    half-life limits \cite{gerda}. The
    expected 90$\%$ CL exclusion reach from GERDA Phase-II \cite{gerdafuture}
    is also shown.} 
    \label{fig:scan1}
\end{figure}
In Fig. \ref{fig:Meejj}, the $M_{eejj}$ distribution is compared with
data~\cite{Khachatryan:2014dka} 
for the background and an example signal model point prediction. We see that
the 
signal is concentrated in the 1.8 TeV$< M_{eejj} <$ 2.2 TeV bin, because the 
width of the slepton is very narrow.
Fig.~\ref{fig:scan1} shows the
$\lambda'_{111}-m_{\tilde{\chi}^0_1}$ plane for {\bf
  S1}-{\bf S3}, each corresponding to a 
different hierarchy of mass parameters $M_1$, $M_2$ and $\mu$. 
It is evident that  a large $\lambda'_{111}$ 
value $\lambda'_{111} \sim 0.4$ is ruled out by the CMS di-jet search
\cite{Chatrchyan:2013qha}.  
In the 1.8 TeV$< M_{eejj} <$ 2.2 TeV bin, CMS measured 1 same sign lepton
pair and 13 opposite-sign pairs. For a given scenario, the ratio $R$ in the
signal 
 of the opposite 
sign to same sign di-leptons ($R$) is predicted to be independent of
$\lambda'_{111}$ and $m_{\tilde{\chi}_1^0}$ to a good approximation. 
{\bf S1} and {\bf S2} predict $R=1.0$ whereas {\bf S3} predicts
$R=3.0$. 
It is difficult for us to estimate whether or not this is a good fit
because we do not know the background rates for same-sign versus opposite sign
leptons. 
We also show the present bound from combined experiments' constraints on the
$0\nu \beta \beta$ decay rate in the figure. 
The region  between the two light curves 
fits the CMS excess at the 95$\%$ CL level. For scenario  {\bf S1}, most of this
`good-fit region' can 
be covered by GERDA Phase-II \cite{gerdafuture}. 
% For scenarios {\bf S1} and {\bf
%   S4}, the constraint from 
% $0\nu \beta \beta$ can be more stringent than the CMS di-jet bound.  
For 
scenario {\bf S2} , a positive signal  in GERDA Phase-II 
is possible in the good-fit region for 
lower neutralino masses $m_{\tilde{\chi}^0_1}<550$ GeV.  However, In {\bf S3},  the
expected reach of GERDA Phase-II does not probe the good-fit region.
{ Note that, although ATLAS and CMS have published searches for like-sign dileptons at 8
TeV in the 20 fb$^{-1}$ data set, those existing in the literature
are not sensitive to our signal, either through large minimum missing transverse momentum  cuts in
${\mathcal R-}$preserving supersymmetry, or through the required presence of
$b-$jets, which our model does not predict.}

To summarize, our model provides a good fit to  the CMS $W_R$ search $eejj$
excess while respecting other empirical constraints. Our model predicts a 
$0\nu\beta\beta$ rate.
Up and coming
$0\nu\beta\beta$ experiments such as GERDA Phase-II will probe a significant
portion of the good-fit parameter space. We look forward to ATLAS providing a
similar analysis of the 8 TeV data, as well as future tests of the excess at
LHC Run II.
We note that our signal comes from 
${\tilde 
    e}_L$ and ${\tilde \nu}$ resonances. With more statistics, it should be
  possible to resolve these 
  two narrow resonances, providing discrimination between our model and others
  that explain the $W_R$ search excess.

{\bf Acknowledgements:} This work has been partially supported by STFC. 
MM and SM  would like to thank the organizers of 
the workshop WHEPP13, Puri, India where part of the work started. 
MM  acknowledges  partial support of the ITN INVISIBLES 
(Marie Curie Actions, PITN-GA-2011-289442). MM would like to thank W. Rodejohann for useful correspondence. SB would like to thank Pradipta Ghosh
for discussions and help in preparing the draft.
We also thank J Chou and J Pastika for helpful communications regarding the
CMS $W_R$ analysis. SM thanks DST, India, for a senior research fellowship.

\end{document}